\documentstyle[]{mn}
\newif\ifAMStwofonts

\def\simgt{\mathrel{\spose{\lower 3pt\hbox{$\sim$}}
        \raise 2.0pt\hbox{$>$}}}
\def\simlt{\mathrel{\spose{\lower 3pt\hbox{$\sim$}}\raise 2.0pt\hbox{$<$}}}

\ifoldfss
  \ifCUPmtlplainloaded \else
    \NewTextAlphabet{textbfit} {cmbxti10} {}
    \NewTextAlphabet{textbfss} {cmssbx10} {}
    \NewMathAlphabet{mathbfit} {cmbxti10} {} 
    \NewMathAlphabet{mathbfss} {cmssbx10} {} 
  \fi
  \ifAMStwofonts
    \ifCUPmtlplainloaded \else
      \NewSymbolFont{upmath} {eurm10}
      \NewSymbolFont{AMSa} {msam10}
      \NewMathSymbol{\upi}     {0}{upmath}{19}
      \NewMathSymbol{\umu}     {0}{upmath}{16}
      \NewMathSymbol{\upartial}{0}{upmath}{40}
      \NewMathSymbol{\leqslant}{3}{AMSa}{36}
      \NewMathSymbol{\geqslant}{3}{AMSa}{3E}

      \let\leq=\leqslant 
       
    \fi
  \fi
\fi 

\ifnfssone
  \newmathalphabet{\mathit}
  \addtoversion{normal}{\mathit}{cmr}{m}{it}
  \addtoversion{bold}{\mathit}{cmr}{bx}{it}
  \newmathalphabet{\mathbfit} 
  \addtoversion{normal}{\mathbfit}{cmr}{bx}{it}
  \addtoversion{bold}{\mathbfit}{cmr}{bx}{it}
  \newmathalphabet{\mathbfss} 
  \addtoversion{normal}{\mathbfss}{cmss}{bx}{n}
  \addtoversion{bold}{\mathbfss}{cmss}{bx}{n}
  \ifAMStwofonts
    \ifCUPmtlplainloaded \else
      \UseAMStwoboldmath
      \makeatletter
      \new@mathgroup\upmath@group
      \define@mathgroup\mv@normal\upmath@group{eur}{m}{n}
      \define@mathgroup\mv@bold\upmath@group{eur}{b}{n}
      \edef\UPM{\hexnumber\upmath@group}
      \new@mathgroup\amsa@group
      \define@mathgroup\mv@normal\amsa@group{msa}{m}{n}
      \define@mathgroup\mv@bold\amsa@group{msa}{m}{n}
      \edef\AMSa{\hexnumber\amsa@group}
      \makeatother
      \mathchardef\upi="0\UPM19
      \mathchardef\umu="0\UPM16
      \mathchardef\upartial="0\UPM40
      \mathchardef\leqslant="3\AMSa36
      \mathchardef\geqslant="3\AMSa3E

      \let\leq=\leqslant 

    \fi
  \fi
\fi 

\ifnfsstwo
  \DeclareMathAlphabet{\mathbfit}{OT1}{cmr}{bx}{it}
  \SetMathAlphabet\mathbfit{bold}{OT1}{cmr}{bx}{it}
  \DeclareMathAlphabet{\mathbfss}{OT1}{cmss}{bx}{n}
  \SetMathAlphabet\mathbfss{bold}{OT1}{cmss}{bx}{n}
  \ifAMStwofonts
    \ifCUPmtlplainloaded \else
      \DeclareSymbolFont{UPM}{U}{eur}{m}{n}
      \SetSymbolFont{UPM}{bold}{U}{eur}{b}{n}
      \DeclareSymbolFont{AMSa}{U}{msa}{m}{n}
      \DeclareMathSymbol{\upi}{0}{UPM}{"19}
      \DeclareMathSymbol{\umu}{0}{UPM}{"16}
      \DeclareMathSymbol{\upartial}{0}{UPM}{"40}
      \DeclareMathSymbol{\leqslant}{3}{AMSa}{"36}
      \DeclareMathSymbol{\geqslant}{3}{AMSa}{"3E}

      \let\leq=\leqslant 

    \fi
  \fi
\fi 

\ifCUPmtlplainloaded \else
  \ifAMStwofonts \else 
    \def\upi{\pi}
    \def\umu{\mu}
    \def\upartial{\partial}
  \fi
\fi

\title[Properties of High Redshift Quasars]{Properties of High Redshift Quasars-I: Evolution of the super-massive black-hole to halo mass ratio}

\author[Wyithe \& Padmanabhan]{J. Stuart B. Wyithe$^{1}$, T. Padmanabhan$^{2}$\\
$^1$ School of Physics, University of Melbourne, Parkville, Victoria, Australia\\
$^2$ Inter-University Center for Astronomy and Astrophysics, Pune, India\\
 Email: swyithe@isis.ph.unimelb.edu.au, nabhan@iucaa.ernet.in}

\date{Accepted Received}
\pagerange{\pageref{firstpage}--\pageref{lastpage}}
\pubyear{2005}

\def\LaTeX{L\kern-.36em\raise.3ex\hbox{a}\kern-.15em
    T\kern-.1667em\lower.7ex\hbox{E}\kern-.125emX}

\begin{document}

\label{firstpage}

\maketitle

\begin{abstract}
\noindent 

In the local universe, the masses of Super-Massive Black-Holes (SMBH)
appear to correlate with the physical properties of their hosts,
including the mass of the dark-matter halos. At higher redshifts, we
observe the growth of SMBHs indirectly through the identification of
high redshift quasars. However information on their hosts is more
difficult to obtain. In this paper we determine the masses of the
halos that host the high redshift quasars (at $z>4$) by comparing the
rate of growth of quasar density with that predicted by the
Press-Schechter mass function. The host mass determined depends on how
the ratio between SMBH and host halo mass evolves with redshift. Under
the assumption that the ratio between SMBH and halo mass does not
evolve with redshift, we find a host halo mass of
$M=10^{11.7\pm0.3}$. Even if the quasars shine at their Eddington
limit, this host mass is significantly smaller than that seen at lower
redshifts in the local universe. Indeed we find that the
null-hypothesis, of a constant ratio between SMBH and halo mass at all
redshifts, can be ruled out at greater than a 5-sigma level. SMBHs
must therefore have contributed a larger fraction to the host mass in
the past. This finding is consistent with expectations from models of
self limiting SMBH growth. When we include the redshift evolution of
the ratio between SMBH and halo mass, we find larger halo masses of
$M\sim10^{12.4\pm0.3}$, in combination with a ratio between SMBH and
host halo mass that increases with redshift in proportion to
$\sim(1+z)^{1.5}$ are required to be consistent with both local and
high redshift observations. We also investigate the restrictions
placed on the critical linear overdensity of quasar hosts at their
epoch of virialisation and find that it cannot exceed the traditional
value of $\delta_{\rm c}=1.69$ by more than a factor of two. Finally,
we find that the high redshift quasars are hosted by fluctuations on
scales that have a variance of $(\delta M/M)=2-3$, corresponding to
$(3-4.5)$-sigma fluctuations in the density field.
\end{abstract}


\section{Introduction}

The Sloan Digital Sky Survey has discovered luminous quasars at
redshifts as high as $z\sim6.4$, i.e., when the universe was only a
billion years old. The super-massive black-holes (SMBH) powering these
quasars have been estimated to have a mass of about $10^9$ solar
masses. However questions regarding the galaxies that host these high
redshift quasars have remained largely unanswered.  To understand the
formation and evolution of quasars and the super-massive black-holes
that power them, one needs to determine several important physical
parameters (such as the quasar lifetime, the ratio of black-hole mass
to halo mass, and the efficiency and rate of accretion during the
luminous phase), as well as the evolution of these parameters with
time. Attempts to answer these questions generally consider the quasar
luminosity function, as it provides a tracer of the density of quasars
with different luminosities as function of cosmic epoch
(e.g. Haehnelt, Natarajan \& Rees 1998; Haiman \& Loeb~1998; Kauffmann
\& Haehnelt~2000; Volonteri, Haardt \& Madau~2003; Wyithe \&
Loeb~2003). However, all these analyses are model dependent and
implicitly assume either (a) the quasar lifetime and its evolution
with redshift, and/or (b) the form, normalisation and evolution of a
relation between SMBH mass and the characteristic velocity of the host
galaxy. The exceptions are studies at low redshift that utilise the
quasar two-point correlation function (Martini \& Weinberg~2001;
Haiman \& Hui~2001; Croom et al.~2004).  At high redshift dynamical
estimates have been made in a few individual cases (Barkana \&
Loeb~2003; Bertoldi et al.~2003).

Locally, direct estimates of SMBH and host mass can be made through
observations of galaxy dynamics.  These observations reveal a
correlation between SMBH mass and the characteristic velocity of the
surrounding stellar spheroid (e.g. Merritt \& Ferrarese~2001; Tremaine
et al.~2002), and by extension of the host dark matter halo
(Ferrarese~2003). These characteristic velocities determine the
dynamical mass, so that there are also corresponding correlations
between SMBH and host mass. Any proposed scenario for SMBH evolution
must reproduce this behaviour and hence the correlations provide
important clues regarding the physics of formation of SMBHs.

However at higher redshift, objects collapse out of a denser
back-ground. The characteristic velocity of a virialised object of a
given halo mass is therefore larger if it formed at higher
redshift. This begs the question of whether the fundamental
correlation is between SMBH mass and the hosts characteristic
velocity, or between SMBH mass and dynamical host mass. Obviously
knowledge of which correlation is fundamental is critical for our
understanding of the astrophysics of SMBH evolution. Unfortunately
since these dynamical observations can only be made for relatively
nearby galaxies, this question cannot be resolved via direct
observation.

Here, within the paradigm of standard concordance cosmology, we show
that by associating halo mass to quasar luminosity, the formation rate
of luminous quasars in the high redshift universe can be used to
constrain the mass of the dark matter halos that host them. We thus
provide a framework within which one can attempt to answer the
question of whether host mass or velocity is the determining factor in
the evolution of a SMBH.  We believe this procedure holds significant
promise for the future when the observations improve.

In \S~\ref{qdensity} we begin with the null-hypothesis that the ratio
between the masses of SMBH and the halo does not evolve with
redshift. Under this assumption, we find the host mass to be about
$10^{11.7\pm0.3}M_\odot$. Estimates for SMBH mass powering the
luminous SDSS quasars yield $\sim10^9M_\odot$ assuming output at the
Eddington limit (the Eddington limit provides a lower limit on SMBH
mass, and hence a lower limit on the mass ratio). The resulting SMBH
to halo mass ratio is therefore much larger for the SDSS quasars than
for local galaxies. Indeed, the null hypothesis can be rejected at a
significance greater 5-sigma, i.e. SMBHs in the past contributed a
larger fraction of galaxy mass than those today. This is one main
conclusion of the paper. Next, we allow for the SMBH to halo mass
ratio to evolve with redshift (\S~\ref{evolve}). In this more general
case we estimate a halo mass for high redshift quasar hosts,
$M\sim10^{12.4\pm0.3}$, and find that the ratio between SMBH and host
halo mass should increase with redshift as $\sim(1+z)^{1.5}$ in order to
be consistent with extrapolation from local observations. Estimates
for the high redshift quasar lifetime are discussed in \S~\ref{lt}. We
also investigate variance of the linear power-spectrum on the scale of
density fluctuations corresponding to the high redshift quasar hosts
in \S~\ref{OMEGA}.  Finally, we discuss the the implications of the
rate of high redshift quasar density evolution for the value of the
critical linear overdensity at host virialisation in \S~\ref{deltac}.
Some concluding discussion is given in \S~\ref{disc}.

Throughout the paper we adopt the set of cosmological parameters
determined by the {\em Wilkinson Microwave Anisotropy Probe} (WMAP,
Spergel et al. 2003), namely mass density parameters of
$\Omega_{m}=0.27$ in matter, $\Omega_{b}=0.044$ in baryons,
$\Omega_\Lambda=0.73$ in a cosmological constant, and a Hubble
constant of $H_0=71~{\rm km\,s^{-1}\,Mpc^{-1}}$. For the primordial
power-spectrum of density fluctuations, we adopt a power-law slope $n
= 1$, and the fitting formula to the exact transfer function of
Bardeen et al.~(1986). It turns out that our results are most sensitive to the cosmological parameter $\sigma_8$, which is the amplitude  of the linearly extrapolated power-spectrum
on scales of $8h^{-1}$Mpc. We present results for different $\sigma_8$ wherever appropriate to illustrate the range of this dependence. 

\section{Evolution in Quasar Density}
\label{qdensity}

\begin{figure*}
\vspace*{65mm}
\includegraphics{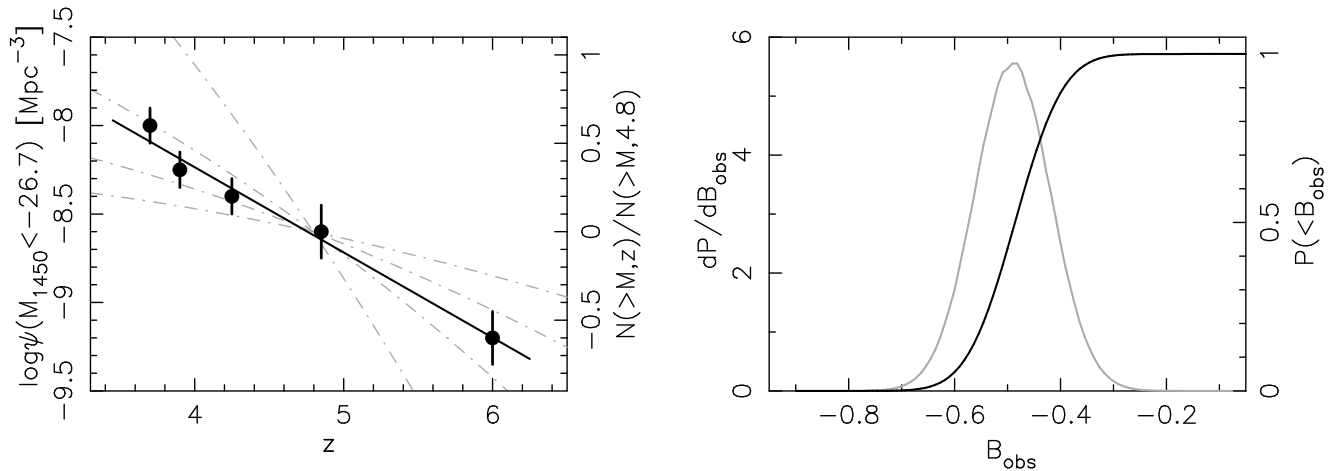}
\caption{Constraints on the evolution of the density of high-redshift quasars. 
{\em Left}: The density of quasars with $M_{1450}<-26.7$ as a
function of redshift (Fan et al.~2001;2003;2004). The solid
line shows the best fit exponential decline
$\Psi(<26.7,z)\propto10^{B_{\rm obs}\times z}$ to the full data set,
with $B_{\rm obs}= -0.49$.  For comparison the evolution in the
density of halos (normalised to unity at $z=4.8$) with masses of
$10^{10}M_\odot$, $10^{11}M_\odot$, $10^{12}M_\odot$ and
$10^{13}M_\odot$ are shown by the grey dot-dashed curves in order of
decreasing steepness. {\em Right}: The corresponding a-posteriori
differential (grey lines; left axis) and cumulative (black lines;
right axis) probability distributions for $B_{\rm obs}$.}
\label{fig1} 
\end{figure*}

The Press-Schechter~(1974) mass function (with the modification of
Sheth \& Tormen~(2002) that will be adopted throughout our
discussion) yields the number density $N(>M(z),z)$ of dark
matter halos above some mass $M(z)$ at redshift $z$. If luminous
quasars reside in a fraction $\epsilon$ of such dark-matter halos,
then the observed number density of quasars is given by the product of
two factors: $N(>M(z),z)$ and $\tau \equiv \epsilon \ {\rm
min}\{t_{\rm q}/H^{-1}(z),1\}$ where $t_{\rm q}$ is the (unknown)
quasar lifetime and $H^{-1}(z)$ is the Hubble time (see e.g., Estathiou \&
Rees~1998).  As a measure of the rate at which luminous quasars
appear, we use the logarithmic derivative $(B)$ of $\tau N(>M(z),z)$,
defined as
\begin{equation}
\label{PSslope}
B=\frac{d\log\tau}{dz} + \frac{\partial\log {N(>M,z)}}{\partial z}
+ \frac{\gamma}{(1+z)}\frac{d\ln N(>M,z)}{d\ln M} 
\end{equation}
Here we have assumed that the halo mass varies as $M\propto(1+z)^\gamma$ at fixed luminosity $M_{1450}$. This choice is convenient for our discussion because observations do
reveal an exponential decline in the quasar population with redshift suggesting approximate constancy of $B$. Our
analysis, of course, relies on the applicability of the Press-Schechter mass
function (as modified by Sheth and Torman~2002).  It has been shown
that this analytic formalism provides an excellent description of the
halo mass function found from numerical simulations (Jenkins et
al.~2001). In particular, if the numerical mass function is expressed
in the appropriate variables then it is independent of epoch, which is
a defining feature of the Press-Schechter formalism. Evaluation of $B$
using the Press-Schechter mass function therefore provides an accurate
description of the redshift evolution of massive dark-matter halos in
a $\Lambda CDM$ cosmology.

The mass function $N(>M,z)$ leads to a steeper slope $B$ as $M$ is
increased. This is illustrated by the grey dot-dashed lines in the
left panel of figure~\ref{fig1} which show the evolution of $N(>M,z)$
(arbitrarily normalised at $z=4.8$) for masses of $10^{10}M_\odot$,
$10^{11}M_\odot$, $10^{12}M_\odot$ and $10^{13}M_\odot$. These curves
should be compared to the observed evolution in the density of
luminous ($M_{1450}<-26.7)$ quasars between redshifts of $z\sim3.7$
and $z\sim6$ from the Sloan Digital Sky Survey (Fan et
al.~2001;2003;2004), which is also summarised in the same panel.  Here
we are assuming that the identification of luminosity $M_{1450}$ with
halo mass $M$ does not vary (that is, $\gamma=0$). This forms the null
hypothesis in this work. The effects of relaxing this assumption are
discussed later in \S~\ref{evolve}.

The curves corresponding to $10^{11}M_\odot$ and $10^{12}M_\odot$ have
logarithmic slopes that lie at the extremes of the range allowed by
the data. \emph{It is therefore clear that if we can measure the
exponential slope in the formation of high redshift quasars, then we
can determine the mass of their host dark-matter halos.}  By using
only the logarithmic slope $B$, we have removed the dependence on the
\textit{absolute value} of quasar lifetime. However, $B$ does depend
on the form of the redshift \emph{evolution} of $\tau$ (and on
$\gamma$). While
$\epsilon$ could also change with $z$ due to various effects (dust
obscuration, beaming angle etc.), we expect the dominant additional
contribution to $B$ to come from $t_{\rm q}$. The $z$ dependence of $t_q$ can be handled
by using two
physically motivated forms for the evolution --- which bracket the
reasonable range of possibilities --- and can be parameterised by
$\tau\propto(1+z)^\alpha$ with $0\la\alpha\la3/2$.  First,
if the quasar lifetime is determined by the mass e-fold timescale of
the SMBHs, then $t_{\rm q}$ is independent of redshift, $\tau\propto
1/H^{-1}(z)$ and $\alpha\approx3/2$.  Second, if the quasar lifetime
is determined by the dynamical timescale at $z$, then $t_{\rm
q}\approx H^{-1}(z)$ making $\tau$ independent of redshift and
$\alpha\simeq0$. (This is also true if $t_{\rm q}> H^{-1}(z)$.)  

\subsection{Evolution constraints}

The evolution shown in figure~\ref{fig1} is well fitted by an
exponential decline (Fan et al.~2001) of the form
\begin{equation}
\label{LF}
\Psi(M_{1450}<-26.7,z)\propto 10^{B_{\rm obs}\times z}.
\end{equation}
The right hand panel of figure~\ref{fig1} shows the a-posteriori
differential (grey curves) and cumulative (black curves) probability
distributions for the observed exponential slope $B_{\rm
obs}$. These distributions were computed as follows. For each value of
$B_{\rm obs}$, we find the normalisation which maximises the product
of probabilities from each redshift bin (assuming Gaussian error
bars). This product represents the likelihood for $B_{\rm obs}$. A
flat prior probability for $B_{\rm obs}$ was then assumed, allowing
calculation of a-posteriori distributions for $B_{\rm
obs}$. Distributions were estimated using $(i)$ the whole data set and
two subsets of the data; $(ii)$ data with $z<5$, and $(iii)$ data with
$z>4.5$. We find that $B_{\rm obs}\sim-0.49\pm0.07$, $B_{\rm
obs}\sim-0.52\pm0.15$ and $B_{\rm obs}\sim-0.53\pm0.20$ describe the
evolution within the full data set and two subsets respectively
showing internal consistency. The best fit is also
plotted in the top left panel to guide the eye.

\subsection{Mass constraints}

\begin{figure*}
\vspace*{70mm}
\includegraphics{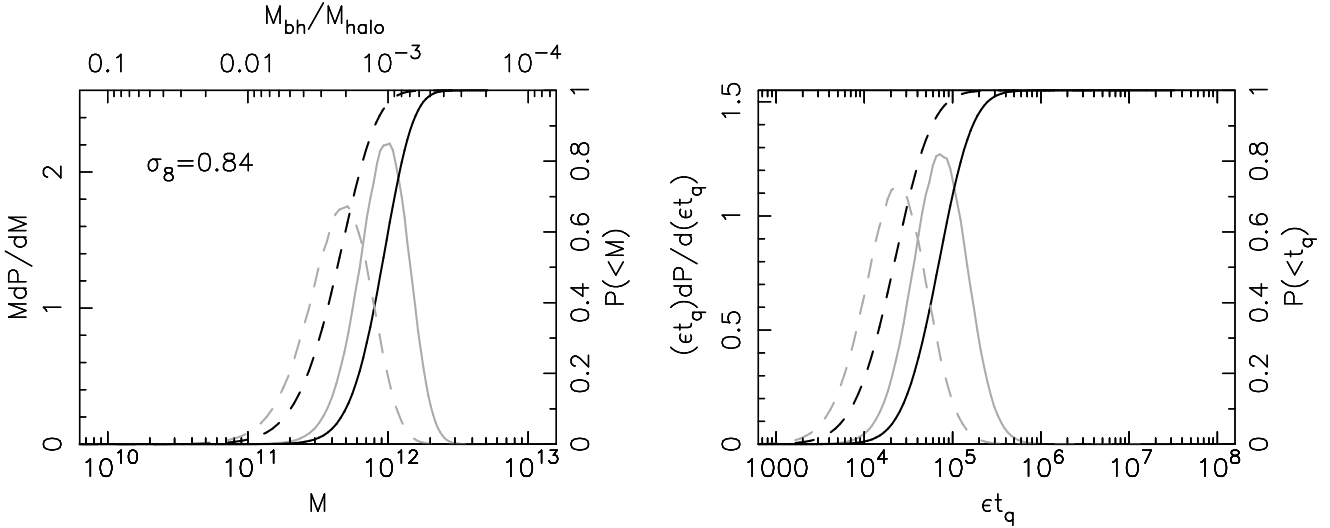}
\caption{Constraints on the mass of the halo that hosts a high redshift
quasar. {\em Left}: The a-posteriori differential (grey lines) and
cumulative (dark lines) probability distributions for $M$ (solid lines
$\alpha=3/2$, dashed lines $\alpha=0$; $\gamma=0$). The upper axis
shows the corresponding values for fraction of halo mass contributed
by a $10^9M_\odot$ black-hole.  {\em Right}: The corresponding
differential (grey lines) and cumulative (dark lines) probability
distributions for $\epsilon t_{\rm q}$. All curves in this figure were
evaluated for $\sigma_8=0.84$.}
\label{fig2} 
\end{figure*}

From equation~(\ref{PSslope}) there is a one-to-one monotonic
correspondence between halo mass $M$ and exponential slope $B$.  The
a-posteriori probability distributions for host halo mass $M$ may
therefore be found by noting that
\begin{equation}
\label{dpdm}
\frac{dP}{dM}\propto\left.\frac{dB}{dM}\right|_{B=B_{\rm obs}}\frac{dP}{dB_{\rm obs}},
\end{equation}
where $(dB/dM)$ was computed using equation~(\ref{PSslope}) at the
central redshift within the luminosity function data. The differential
(grey lines) and cumulative (dark lines) a-posteriori distributions
for $M$ are plotted in the left hand panel of figure~\ref{fig2}. If
$\alpha=3/2$ (solid lines), the observed evolution in the density of
bright quasars implies host halos with masses of
$10^{11.9\pm0.2}M_\odot$.  If $\alpha=0$ (dashed lines), then the halo
mass estimates and their upper bounds are about a factor of $2$
smaller; we get $10^{11.6\pm0.2}M_\odot$.  Thus under the
null-hypothesis of a fixed $\gamma=0$, the exponential slope of the
high redshift quasar luminosity function leads to the determination of
masses of halos that host high redshift quasars to within a factor of
a few.  In particular, we stress that the result does not rely on any
a-priori assumptions about the relation between quasar luminosity and
SMBH mass, about the relation between SMBH and halo mass, or about the
halo density profile.

As an aside, we
note that an alternative approach to equations~(\ref{dpdm}) [and
(\ref{dPdlt})] for calculation of $dP/dM$ [and $dP/d(\epsilon t_{\rm
q})$] would be to choose prior probabilities for $M$ and $\epsilon
t_{\rm q}$ that are flat in the logarithm, and likelihoods for $M$ and
$\epsilon t_{\rm q}$ based on the comparison of the corresponding
slope with the data. We find that this approach gives nearly identical
results to those presented in this paper.

\subsection{SMBH to halo mass ratio}

We now explore several further consequences.  For the estimation of
SMBH mass, it is usually assumed that quasar emission is isotropic,
and that emission is at the Eddington rate, resulting in SMBHs
powering the highest redshift quasars having inferred masses (Fan et
al.~2001) of $\sim10^{9}M_\odot$. These estimates are consistent with
dynamical estimates based on emission line profiles (Willott, McLure
\& Jarvis~2003). We have therefore labeled the upper axis of the left
panel in figure~\ref{fig2} with the fraction of halo mass contributed
by a $10^9M_\odot$ black-hole, allowing the curves in this panel to
represent the a-posteriori probability distributions for this fraction
as well as $M$. We find that the full data set implies SMBHs
contribute a fraction of about $10^{-2.9\pm0.2}(M_{\rm
bh}/10^9M_\odot)$ and $\sim10^{-2.6\pm0.2}(M_{\rm bh}/10^9M_\odot)$ of
the halo mass for $\alpha=3/2$ and $\alpha=0$ respectively.  These
fractions are larger than those found by Ferrarese~(2003) for
\textit{local} $\sim10^{12}M_\odot$ galaxies, which are $M_{\rm
bh}/M=10^{-5.6}$, $10^{-5}$ and $10^{-4.2}$ respectively under the
assumptions of singular isothermal halos, Navarro, Frenk \&
White~(1997) (NFW) halos, and halo masses derived from galaxy-galaxy
lensing (Seljak~2002).  

This result has two possible interpretations. First, the SMBHs are
accreting at well above their Eddington rate and/or with high
efficiency so that the black-hole mass has been overestimated (this
disagrees with dynamical studies, Willott et al.~2003). A second, more
agreeable, interpretation is that SMBHs may contribute a larger
fraction of the halo mass at higher redshifts. Indeed this latter
scenario is independently supported by observations of quasar host
galaxies at $z\sim2$ which suggest black-hole masses that are
significantly larger with respect to their hosts than in the case of
local galaxies (Croom et al.~2004).

\subsection{Sensitivity to $\sigma_8$}

\begin{figure*}
\vspace*{70mm}
\includegraphics{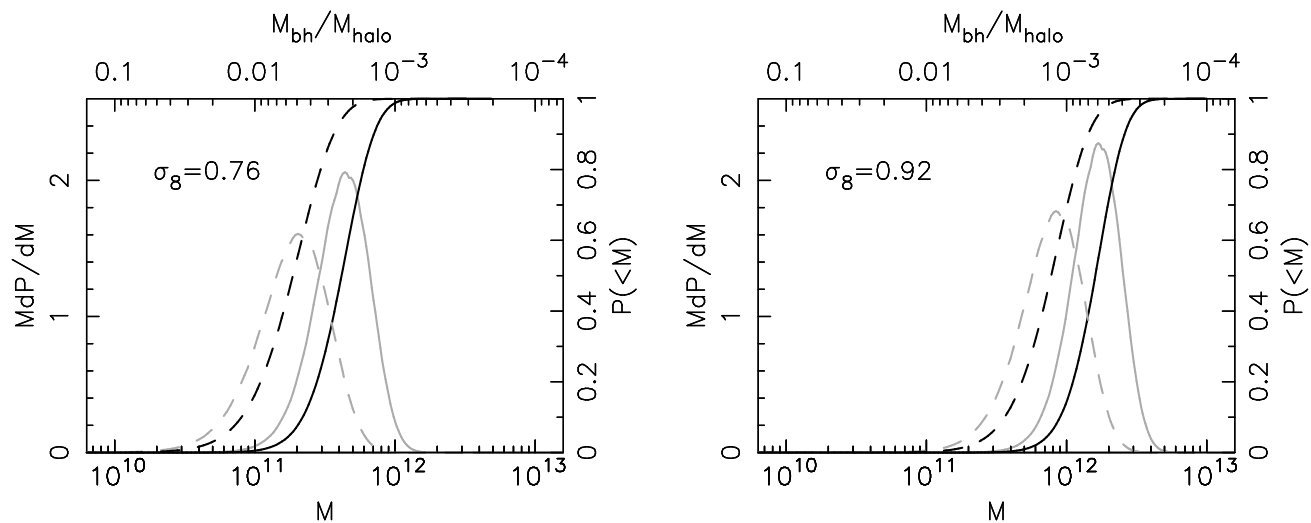}
\caption{Constraints on the mass of dark matter halos 
which host quasars. The a-posteriori differential (grey lines) and
cumulative (dark lines) probability distributions for $M$ obtained
using values of $\sigma_8=0.76$ and $\sigma_8=0.92$. These correspond
to the $2-\sigma$ range for $\sigma_8$ determined from WMAP (Spergel
et al.~2003). The solid and dashed curves in these panels correspond
to $\alpha=3/2$ and $\alpha=0$. In each panel the upper axis shows the
corresponding values for the fraction of halo mass contributed by a
$10^9M_\odot$ black-hole. }
\label{fig3} 
\end{figure*}

Of the observable cosmological parameters, the relationship between
$B$ and $M$ is most sensitive to $\sigma_8$.  To illustrate the extent
of this dependence we have repeated our analysis using values of
$\sigma_8=0.76$ and $\sigma_8$=0.92. (These values bound the
2-$\sigma$ range of the best fit constraints derived from {\em WMAP}
plus large-scale structure and Ly$\alpha$ forest data Spergel et al~2003.) 
We find that the constraints on the quasar host
halo mass vary by a factor of $\sim4$ within the $2-\sigma$ range for
$\sigma_8$ (see figure~\ref{fig3}). 

\subsection{Marginalised distributions}

\begin{figure*}
\vspace*{70mm}
\includegraphics{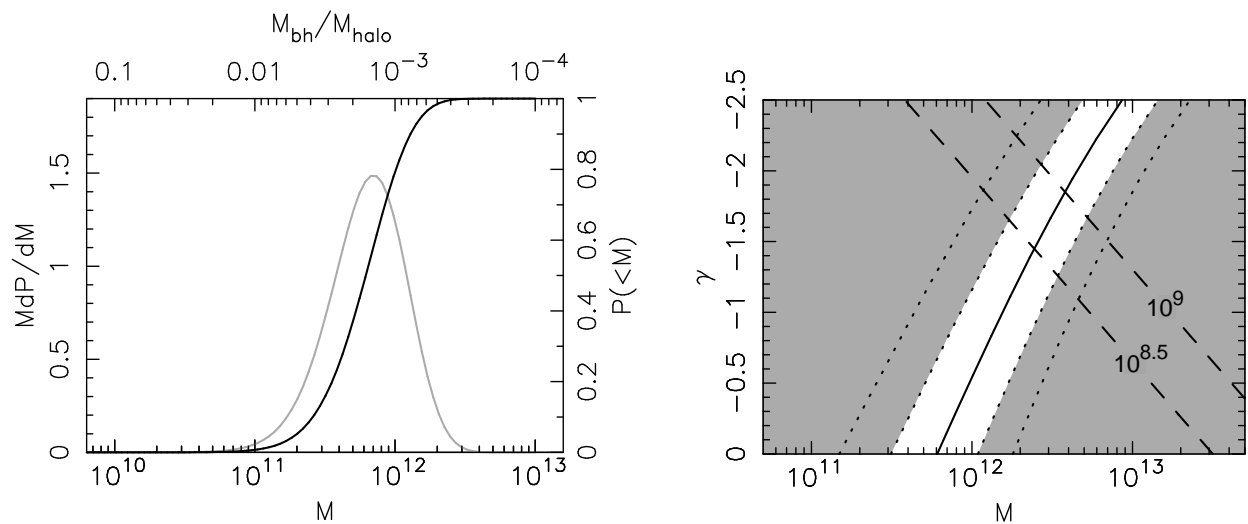}
\caption{Constraints on the mass of dark matter halos 
which host quasars. {\em Left}: A-posteriori probability
distributions for $M$ that have been marginalised over $\alpha$ and
$\sigma_8$ (see equation~\ref{marginalise}). The upper axis shows the
corresponding values for the fraction of halo mass contributed by a
$10^9M_\odot$ black-hole. {\em Right}: The 2.5, 16, 50, 84 and
97.5 percentiles of the cumulative marginalised probability for mass
$M$ as a function of $\gamma$. The grey regions represent masses
outside the 1-$\sigma$ range for each $\gamma$. For comparison, we show
curves (dashed lines) representing quasar host halo mass at high
redshift derived assuming the local black-hole halo mass
ratio (Ferrarese~2003; for a $10^{8.5}M_\odot$ and $10^9M_\odot$ SMBH, and
the NFW profile) plus an evolution of this ratio with redshift that is
proportional to $(1+z)^{-\gamma}$. }
\label{fig4} 
\end{figure*}

One can also marginalise over systematic uncertainty in $dP/dM$
due to $\alpha$ and $\sigma_8$;
\begin{equation}
\label{marginalise}
\frac{dP}{dM}\propto\int_0^{3/2}d\alpha\int_0^\infty d\sigma_8 \exp{\left[\frac{-(\sigma_8-0.84)^2}{2(0.04)^2}\right]}\frac{dP}{dM}(\alpha,\sigma_8), 
\end{equation}
where $(dP/dM)(\alpha,\sigma_8)$ was determined using
equation~(\ref{dpdm}), and we have used a flat prior probability for
$\alpha$ in the range $0\leq\alpha\leq3/2$ combined with a Gaussian
probability distribution for $\sigma_8$. We find (left hand panel of
figure~\ref{fig4}) a value of $M=10^{11.7\pm0.3}M_\odot$, which is our
best estimate for the halo mass of the high redshift galaxies that
host the quasars under the null hypothesis of $\gamma=0$. Assuming a
$10^9M_\odot$ central black-hole, corresponding to accretion at the
Eddington limit (Willott, McLure \& Jarvis~2003), this corresponds to
a black-hole to halo mass ratio of $10^{-2.7\pm0.3}$, which is
inconsistent with local estimates (Ferrarese~2003; $\sim10^{-5}$
assuming an NFW profile) at greater than $7-\sigma$. If luminous
quasars shine near their Eddington rate over a range of redshifts, we
therefore conclude that the SMBH to halo mass ratio must increase with
redshift. 

\subsection{Mass estimates assuming an evolving SMBH to halo mass ratio}
\label{evolve}

\begin{figure*}
\vspace*{70mm}
\includegraphics{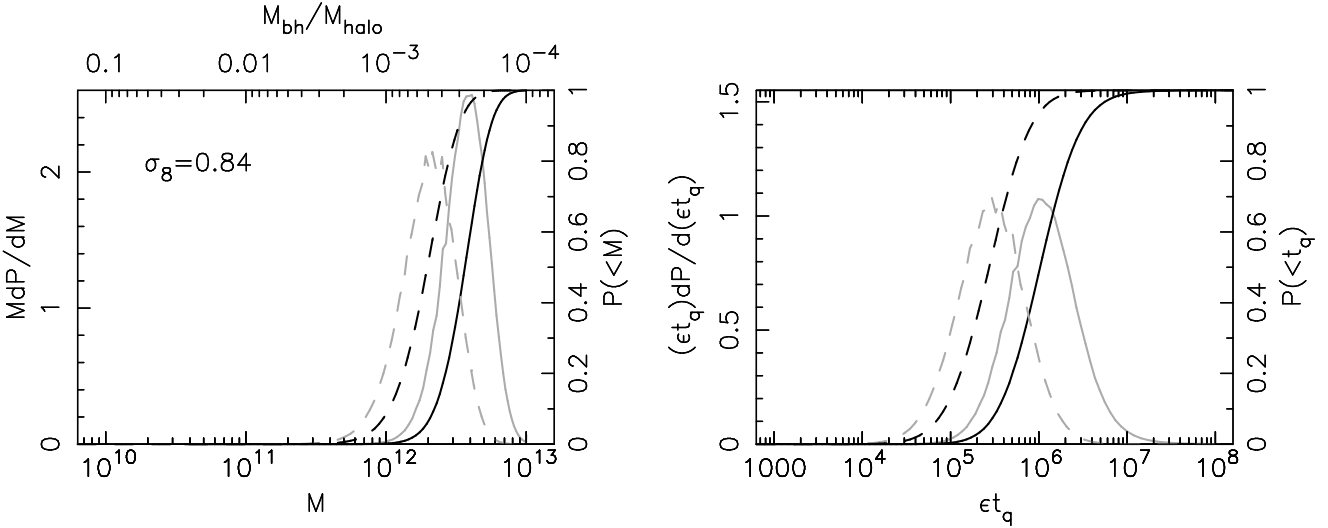}
\caption{Constraints on the mass of the halo that hosts a high redshift
quasar. {\em Left}: The a-posteriori differential (grey lines) and
cumulative (dark lines) probability distributions for $M$ (solid lines
$\alpha=3/2$, dashed lines $\alpha=0$; $\gamma=-1.5$). The upper axis
shows the corresponding values for fraction of halo mass contributed
by a $10^9M_\odot$ black-hole.  {\em Right}: The corresponding
differential (grey lines) and cumulative (dark lines) probability
distributions for $\epsilon t_{\rm q}$. All curves in this figure were
evaluated for $\sigma_8=0.84$.}
\label{fig5} 
\end{figure*}

In evaluating the derivative $B$, we have thus far assumed a
luminosity $M_{1450}$ to be associated with a fixed halo mass
($\gamma=0$). However if the halo mass $M$ housing quasars of
luminosity $M_{1450}$ varies with redshift, then the evaluation of $B$
must include the extra term in equation~(\ref{PSslope}) of the form
$\gamma M(z)/(1+z)\times d\log_{10}N(>M,z)/dM$ (where we have assumed
that the halo mass varies as $M\propto(1+z)^\gamma$ at fixed
luminosity $M_{1450}$.) In the right hand panel of figure~\ref{fig4}
we show the 2.5, 16, 50, 84 and 97.5 percentiles of the cumulative
marginalised distribution (equation~\ref{marginalise}) for halo mass
as a function of $\gamma$. The grey regions represent the masses
outside the 1-$\sigma$ range for each $\gamma$. Smaller values of
$\gamma$ lead to larger estimates of the mass (Note that the y-axis is
reversed with $\gamma$ decreasing from bottom to top). If the
evolution of halo mass $M$ housing a fixed black-hole mass follows
$M\propto(1+z)^\gamma$ between the local and high redshift universe
then we can estimate $M$ at high redshift as a function of $\gamma$,
by extrapolating the local relation (Ferrarese~2003) between halo and
black-hole mass. The resulting curves (taking the case of an NFW
profile) are plotted for black-hole masses of $10^{8.5}M_\odot$ and
$10^9M_\odot$ (dashed lines from left to right). By comparing the two
constraints we find $\gamma\sim-(1.5-2)$, leading to estimates of halo
mass that are $\sim3-6$ times larger than our evolution-free
($\gamma=0$) estimate. Note that these larger masses do not weaken our
result that SMBHs comprised a larger fraction of galaxy mass at high
redshift, as this behaviour is explicit when
$\gamma<0$. Interestingly, values of $\gamma<0$ follow naturally from
models where SMBH growth is self limiting through feedback on galactic
gas (e.g. Haehnelt, Natarajan \& Rees~1998; Wyithe \& Loeb~2003).

We see from Figure~\ref{fig4} that if the evolution of the ratio
between SMBH and halo mass (or more correctly, the ratio between
quasar luminosity and halo mass) can be described as a powerlaw in
redshift, then an extrapolation from local observations combined with
our evolution analysis implies a value of $\gamma\sim-3/2$. In the left
panel of Figure~\ref{fig5} we plot the differential and cumulative
probability distributions for $M$ assuming $\gamma=-3/2$. As mentioned
in the previous paragraph we find larger masses than were derived in
Figure~\ref{fig2} under the assumption of non-evolving ratio; we get
$M=10^{12.5\pm0.2}M_\odot$ ($\alpha=3/2$) and
$M=10^{12.3\pm0.2}M_\odot$ ($\alpha=0$).

\section{quasar lifetime}
\label{lt}

If a fraction $\epsilon$ of dark matter halos contain SMBHs, then the
total lifetime of the quasar can be estimated (Martini \&
Weinberg~2001; Haiman \& Hui~2001) by dividing the quasar number density
$\Psi(M_{1450}<-26.7,z)$ by $\epsilon$ times the number density $N(>M)$ of
halos larger than $M$, and then multiplying by the Hubble time (for
$t_{\rm q} < H^{-1}$).  The a-posteriori probability for the product
$\epsilon t_{\rm q}$ is
\begin{equation}
\label{dPdlt}
\frac{dP}{d(\epsilon t_{\rm q})}\propto \left[\frac{d}{dM}\int d\Psi \frac{dP}{d\Psi}\epsilon t_{\rm q}(\Psi,M)\right]^{-1} \left.\frac{dB}{dM}\right|_{B=B_{\rm obs}} \frac{dP}{dB_{\rm obs}},
\end{equation}
where $(dP/d\Psi)$ is the observed Gaussian probability for $\Psi$.
Under the assumption of a SMBH to halo mass ratio that does not evolve
with redshift (see Figure~\ref{fig2}, right hand panel), we find
lifetimes of $10^{4.8\pm0.3}\epsilon^{-1}$ and
$10^{4.3\pm0.4}\epsilon^{-1}$ years (respectively) for $\alpha=3/2$
and $\alpha=0$, in a cosmology where $\sigma_8=0.84$. In addition (not
shown) we have computed distributions for $\epsilon t_{\rm q}$
corresponding to $\sigma_8=0.76$ and $\sigma_8$=0.92. We find
constraints that vary by a factor of 2 relative to the case of
$\sigma_8=0.84$.  These results imply that if all dark matter halos contained
SMBHs at high redshift ($\epsilon=1$) then the preferred quasar
lifetime of $(10^4-10^5)$ years is significantly shorter than both the
Salpeter time [about $4\times10^7(\epsilon_{\rm eff}/0.1)\eta^{-1}$
years] for accretion at the maximal rate ($\eta=1$) with
$\epsilon_{\rm eff}=10\%$ efficiency of conversion from mass to
energy, as well as estimates of the quasar lifetime at lower redshifts
[($10^6-10^8$) years, see Martini~(2003) for a summary]. The small
value of $\epsilon t_{\rm q}$ might therefore indicate that not all
dark-matter halos at high redshift contain supermassive black-holes,
in contrast to the situation locally (Kormendy \&
Richstone~1995). Indeed, if the high redshift quasar lifetime were
$(10^{6}-10^{8})$ years, as seems to be the case (Yu \& Tremaine~2002)
at $z\sim2$, this would imply that only 1 in $(10-10^3)$ dark matter
halos at $z>4$ contained a SMBH.  Alternatively, there could be a
larger number of obscured quasars at high redshift, giving the
impression of a smaller quasar lifetime.

On the other hand, we have suggested that a situation where the SMBH
to halo mass ratio does not evolve with redshift will be inconsistent
with local observations. As a result it should not be surprising that
the lifetime derived when $\gamma=0$ is inconsistent with other
observations.  In \S~\ref{evolve} we derived results for the halo mass
under the assumption that the SMBH to halo mass ratio evolve as a
power-law in redshift $[\propto(1+z)^{-\gamma}]$ and found that larger
masses of $M\sim10^{12.4\pm0.3}M_\odot$ are obtained for
$\gamma=-1.5$. We have also derived the a-posteriori probability
distribution for quasar lifetime in this case (right panel of
Figure~\ref{fig5}), and found that the larger halo mass leads to a
longer inferred lifetime. We find $t_{\rm
q}=10^{5.9\pm0.4}\epsilon^{-1}$ years ($\alpha=3/2$) and $t_{\rm
q}=10^{5.5\pm0.5}\epsilon^{-1}$ years ($\alpha=0$). For occupation
fractions of unity ($\epsilon=1$) these lifetimes are marginally
consistent with, though still smaller than the lifetime of
$10^{6}-10^8$ years inferred at lower redshift.

\section{The variance of density fluctuations corresponding to high redshift quasar hosts}
\label{OMEGA}

\begin{figure*}
\vspace*{65mm}
\includegraphics{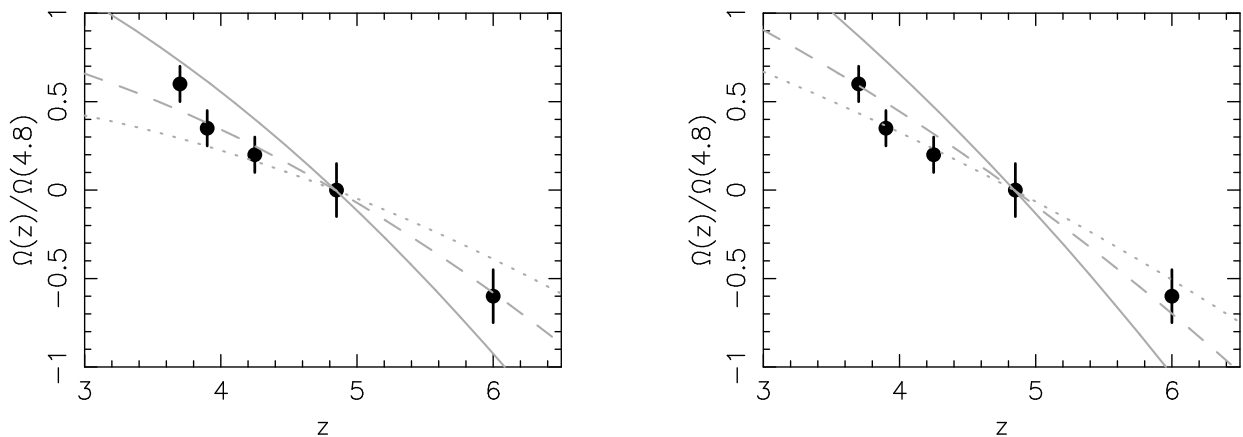}
\caption{The evolution of density with redshift for different values of linear
variance $\sigma$ corresponding to halos that host high redshift
quasars. The curves correspond to $\sigma=2$ (solid line),
$\sigma=2.5$ (dashed line) and $\sigma=3$ (dotted line). Both the
model and quasar densities have been normalised to unity at
$z=4.8$. {\em Left:} curves for $\alpha=3/2$. {\em Right:} curves for
$\alpha=0$.}
\label{fig6} 
\end{figure*}

An alternative way of studying the hosts of the high redshift quasars
is to compare the evolution of their density with the evolution of the
density of halos that correspond to scales with a fixed variance in
the linearly extrapolated power-spectrum. In Figure~\ref{fig6} we show
curves of density [$\tau N(>M(\sigma),z)$] against redshift for values
of variance $\sigma=2$, 2.5 and 3. Since we are interested in the {\it
deviation} of the observed results from the theoretical curves for a
constant $\sigma$, we have normalised both the theoretical curves (from
Press-Schechter) and the data to unity at an intermediate redshift
$z=4.8$, and thus express the evolution in terms of a dimensionless
density parameter [$\Omega(z)/\Omega(4.8)$]. The left and right panels
correspond to values of $\alpha=3/2$ and $\alpha=0$ respectively.  We
find that the values of $\sigma=2-3$ bracket the range of evolution,
implying that for a linearly extrapolated critical overdensity of
$\delta_c\sim 1.69(1+z)$, the high redshift quasar dark-matter host
galaxies formed from $\sim3$-sigma to $\sim4.5$-sigma density
fluctuations. The rareness of these fluctuations is consistent with
the large masses (well in excess of the non-linear mass-scale) which
were inferred in \S~\ref{evolve}.

\section{the linear overdensity for quasar hosts}
\label{deltac}

\begin{figure*}
\vspace*{65mm}
\includegraphics{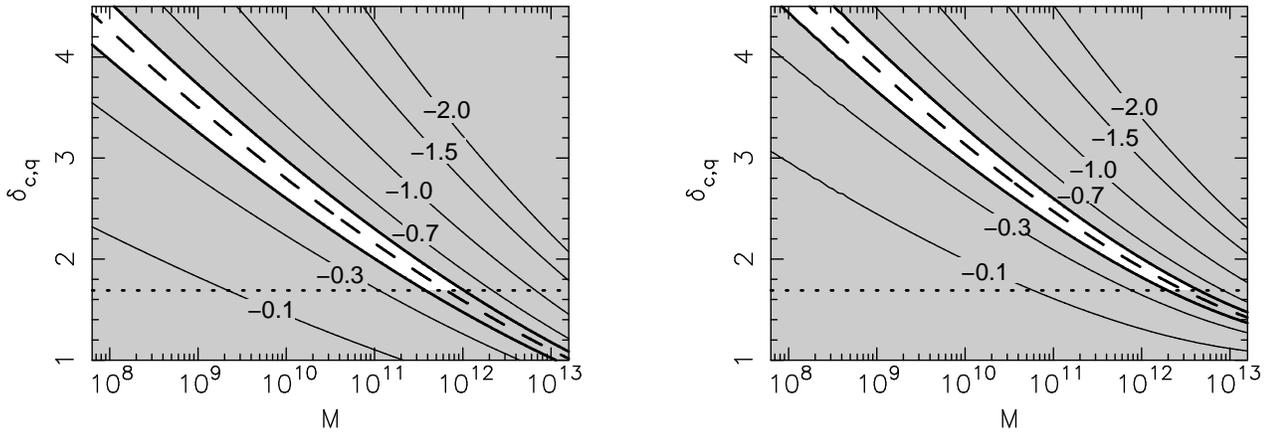}
\caption{Contours of the exponential slope $B$ as a function of the linear
overdensity for quasars $\delta_{\rm c,q}$ and host halo mass $M$ (we
assumed $\alpha=3/2$). The shaded regions are excluded by the
requirement that $\delta_{\rm c,q}>1.69$, and by the limits on $B_{\rm
obs}$ derived from the redshift evolution in the quasar luminosity
function. The dashed line shows the most likely value for the slope,
and the dotted line corresponds to $\delta_{\rm c,q}=1.69$. {\em
Left:} curves for $\gamma=0$. {\em Right:} curves for $\gamma=-1.5$. All
curves in this figure were evaluated for $\sigma_8=0.84$.}
\label{fig7} 
\end{figure*}

When calculating the Press-Schechter mass function, we have so far
adopted the conventional value for the linear overdensity at halo
virialisation of $\delta_{\rm c}=1.69$, which is appropriate for a
spherical collapse at the time when shells evolve to zero radius.
This value of $\delta_{\rm c}$, which corresponds to a non-linear
overdensity of about $178$, agrees with results from N-body
simulations (Jenkins et al~2001).  However, in general, non-spherical
top-hat models virialise at a time when the linear overdensity reaches
a value greater (Engineer, Kanekar \& Padmanabhan~2000) than
$\delta_{\rm c}=1.69$. 

As we showed in \S~\ref{OMEGA}, luminous quasars are rare systems
forming from greater than 3-sigma fluctuations. It is therefore
possible that they form only in hosts of unusual overdensity.  In
figure~\ref{fig7} we have plotted contours of $B$ as a function of
$\delta_{\rm c}$ and $M$.  [for the case of $\alpha=3/2$ and
$\sigma_8=0.84$, and assuming $\gamma=0$ (left panel) and $\gamma=-3/2$
(right panel).] The shaded grey regions show the excluded values of
$\delta_{\rm c}<1.69$, and $B$ (68\% range).  The most likely value of
the variance is plotted as a thick dashed line. We see that
$\delta_{\rm c}$ cannot be too different from 1.69, or else the host
dark matter halos would be unacceptably small. This indicates that
collapse to an unusually overdense halo is not taking place in halos
that host SMBH formation and quasar activity. An absolute upper limit
on $\delta_{\rm c}$ can be obtained by noting that if the mass of the
SMBH is $10^9M_\odot$, then host masses must be larger than about
$10^{10}M_\odot$. In this case, values of $\delta_{\rm c}\ga3$ are not
allowed. If the SMBH were restricted to contain less than 10\% of the
gas component, then $\delta_{\rm c}\la2$.

\section{Discussion}
\label{disc}

In this paper we have estimated the mass of high redshift quasar host
dark matter halos by comparing the rate of quasar evolution with the
Press-Schechter mass function. We have found that in the case of the
null hypothesis where the SMBH to halo mass ratio does not change with
redshift, that the implied halo mass is
$M=10^{11.7\pm0.3}M_\odot$. This mass is significantly smaller than
the mass of local halos that house a $10^9M_\odot$ SMBH, the mass
believed to be powering the SDSS quasars. Indeed our results rule out
the null-hypothesis at a significance greater than 5$-\sigma$. We
therefore conclude that the SMBH to halo mass ratio must increase
towards higher redshift, i.e. SMBHs contained a larger fraction of the
host galaxy mass at earlier times.

Having demonstrated that SMBHs at high redshift must have contained a
larger fraction of the host mass than SMBHs observed today, we allowed
the SMBH to halo mass ratio to vary with redshift. In this case it is
possible to achieve consistency between observations of high redshift
quasars and local SMBH masses. We find that these combined constraints
imply high redshift quasar host halo masses of $M=10^{12.4\pm0.3}$,
with a SMBH to halo mass ratio that varies with redshift approximately
as $(1+z)^{3/2}$.

A scenario where SMBHs formed at high redshift contain a greater
fraction of the host galaxies mass is consistent with models of SMBH
evolution in which SMBH growth is limited by feedback during the
quasar phase (e.g. Silk \& Rees~1998; Haehnelt, Natarajan \&
Rees~1998; Wyithe \& Loeb~2003). These models predict a relation
between SMBH mass and the characteristic velocity of the host which is
redshift independent. As a result feedback regulated growth of SMBHs 
leads naturally to a ratio of SMBH to halo mass that increases with
redshift. Our results therefore support feedback regulated schemes
where SMBH growth is dominated by accretion during the luminous quasar
phase.

Upon completion, the Sloan Digital Sky Survey will have identified
much larger numbers of high redshift quasars than are currently
published . The more accurate luminosity functions which will be available
should then allow a similar and more precise analysis using the
model-independent technique introduced in this work. Such an analysis
may allow the variation of host mass with redshift to be determined
directly for the high redshift quasars.

\section*{Acknowledgements}

The work of JSBW was supported by the Australian
Research Council.  This work was initiated when one of the
authors (TP) was visiting the School of Physics, University of
Melbourne, under the Miegunah Distinguished Fellowship.

\label{lastpage}

\end{document}